\documentclass{iaus}
\usepackage{graphicx}
\usepackage{subfigure}
\usepackage{multirow}

\title[~~3D Structure of NGC 891 and M 51] 
{The 3-Dimensional Structure of NGC 891 and M51}

\author[A. Schechtman-Rook, M. A. Bershady, K. Wood, \and T. P. Robitaille]   
{Andrew Schechtman-Rook$^1$, Matthew A. Bershady$^1$, Kenneth Wood$^{2,1}$, \and Thomas P. Robitaille$^3$}

\affiliation{$^1$Department of Astronomy, University of Wisconsin-Madison,\\ 425 N. Charter St., Madison, WI 53706\\ email: {\tt andrew@astro.wisc.edu\\mab@astro.wisc.edu} \\[\affilskip]
$^2$School of Physics \& Astronomy, University of St Andrews,\\ North Haugh, StAndrews, Fife, Scotland, KY16 9SS\\email: {\tt kw25@st-andrews.ac.uk}\\[\affilskip]
$^3$Harvard-Smithsonian Center for Astrophysics, \\ 60 Garden St., Cambridge, MA 02138\\email: {\tt trobitaille@cfa.harvard.edu}}

\pubyear{2011} 
\volume{284}  
\pagerange{1--12}
\jname{The Spectral Energy Distribution of Galaxies}
\editors{R.J. Tuffs \&  C.C.Popescu, eds.}
\begin{document}

\maketitle

\begin{abstract}
We investigate the three-dimensional structure of the nearby edge-on spiral galaxy NGC 891 using 3D Monte Carlo radiative transfer models, with realistic spiral structure and fractally clumped dust. Using the spiral and clumpiness parameters found from recently completed scattered light models we produce lower resolution SED models which reproduce the global UV-to-FIR SED of NGC 891. Our models contain a color gradient across the major axis of the galaxy - similar to what is seen in images of the NGC 891. With minor adjustment our SED models are able to match the majority of M51’s SED, a similar galaxy at a near face-on different inclination. 
\keywords{radiative transfer, galaxies: ISM, galaxies: spiral, galaxies: structure}
\end{abstract}

\firstsection 
\section{Introduction}
The interplay between starlight and dust in spiral galaxies is enormously complex. In the last decade, space-based observatories have allowed astronomers to measure spectral energy distributions (SEDs) from the ultraviolet (UV) through the far-infrared (FIR). These SEDs show that radiative transfer (RT) models based only on optical data (e.g. \cite[Xilouris et al. 1999]{Xilouris99}) significantly overpredict the amount of escaping UV radiation and underpredict the amount of dust present in edge-on spirals (\cite[Popescu et al. 2000]{Popescu00}; \cite[Bianchi 2008, hereafter B08]{Bianch08}). To combat this problem astronomers have created more complex models, including additional stellar disk components and semi-analytic approximations for the effects of enshrouded star formation with clumpy dust and light. 

Most of these “advanced” models have focused exclusively on fitting galaxy SEDs at the expense of creating models with observationally accurate images, due to the difficulty in fitting non-axisymmetric parameters like clumpy dust substructures or spiral arms. 

We have recently completed a preliminary investigation into optimizing non-axisymmetric RT models fit to high-resolution {\it Hubble Space Telescope} optical images of the central region of the edge-on spiral galaxy NGC 891 (\cite[Schechtman-Rook et al. 2011]{Schechtman-Rook12}). These models included fractally clumped dust and logarithmic spiral arms specially designed to match observations of the arm/interarm ratios of face-on spiral galaxies. The best-fitting models show a strong preference for spirality and clumpiness. Here we investigate the effect of the spirality and clumping parameters determined from these fits to optical images on fitting the SEDs of NGC 891 and M51. We use a new 3D Monte Carlo RT code (HYPERION, \cite[Robitaille et al. 2011]{Robitaille11}), which uses arbitrary dust and emissivity grids and includes multiple dust grain sizes (including PAHs).

\section{Model}
Our 100x100x50 pixel SED models are mapped to a
40x40x20 kpc volume. Stellar emission is set to the Sb spectrum of
\cite[Fioc \& Rocca-Volmerange (1997)]{Fioc97}, split into a old population of
stars with masses $<$3 M$_\odot$(bulge and Disk-1) and an young
population (M $> 3$ M$_\odot$; Disk-2). Model parameters are shown in
Table 1. M$_{\rm d}$ is the total dust mass, L$_1$ and L$_2$ are the
total bolometric {\it disk} luminosities, and T denotes the density
threshold for young stellar emission (where appropriate). All models
have a Sersic n=4 bulge with L$_{\rm bulge}=1.8\times
10^{10}$L$_\odot$ and structural parameters identical to those used in
Bianchi (2008). The model ratios of old-to-young stellar populations span unredenned,
spatially-integrated SEDs corresponding to a wide range of Hubble
types (for reference template E,Sa,Sc,Im galaxies from \cite[Fioc \& Rocca-Volmerange (1997)]{Fioc97} have old-to-young ratios of 35,9.8,2.3,1.1,
respectively).

Disk-1 has half the scale-length and twice the scale-height of the dust.
Disk-2 has the same scales as the dust, and when it is clumpy it
follows the distribution of the clumpy dust; higher dust density cells
weighted to have more stellar emission. Clumpiness and spirality are
consistent with what we found in Schechtman-Rook et al. (2011), i.e.,
50\% of the dust mass is put in clumps distributed in 130 parent
cells; spiral arms have a pitch angle of 20$^\circ$ and a relative
strength of 50\%. For models 4 and 5 there is a minimum dust density
required to place Disk-2 emission in a cell; this threshold mimics the
effects of embedded star formation. 

\begin{table}
\begin{centering}
\caption{Model Parameters}
\vspace{2.5 mm}
\begin{tabular}{cccccccc}
\hline\hline\relax\\[-1.7ex]
 & & Clumpy & Clumpy & M$_{\mathrm{d}}$ & L$_{1}$ & L$_{2} $ &T \\
Model & Spirality & Dust & Disk 2 &($10^{8}$ M$_\odot$) & ($10^{10}$ L$_\odot$) &($10^{10}$ L$_\odot$) & (g cm$^{-3}$)\\
\hline
1 & No & No & No & 1.0 & 4.2 & 0.1 & ---\\
2 & No & Yes & No & 1.0 & 4.2 & 0.8 & ---\\
3 & Yes & Yes & Yes & 1.5 & 4.2 & 0.8 & 0\\
4 & Yes & Yes & Yes & 1.0 & 4.2 & 2.8 & 3.42 $\mathrm{x}\, 10^{-27}$\\
5 & Yes & Yes & Yes & 1.75 & 2.0 & 2.8 & 3.42 $\mathrm{x}\, 10^{-27}$\\
\hline
\end{tabular}
\label{tab1}
\end{centering}
\end{table}
\section{Results}
Single disk models are unable to simultaneously match the UV and FIR SED of NGC 891, even when the dust mass and stellar luminosity are increased or clumpiness and spirality are added. The poor fit from single disk models was also noted by \cite[B08]{Bianchi08}. Adding spirality, regardless of the amplitude of the spiral perturbations, has almost no effect on the spatially-integrated SED morphology, a result found also by \cite[Popescu et al. (2011)]{Popescu11}. 

\begin{figure}
\begin{centering}
 \subfigure{\includegraphics[width=2.44in]{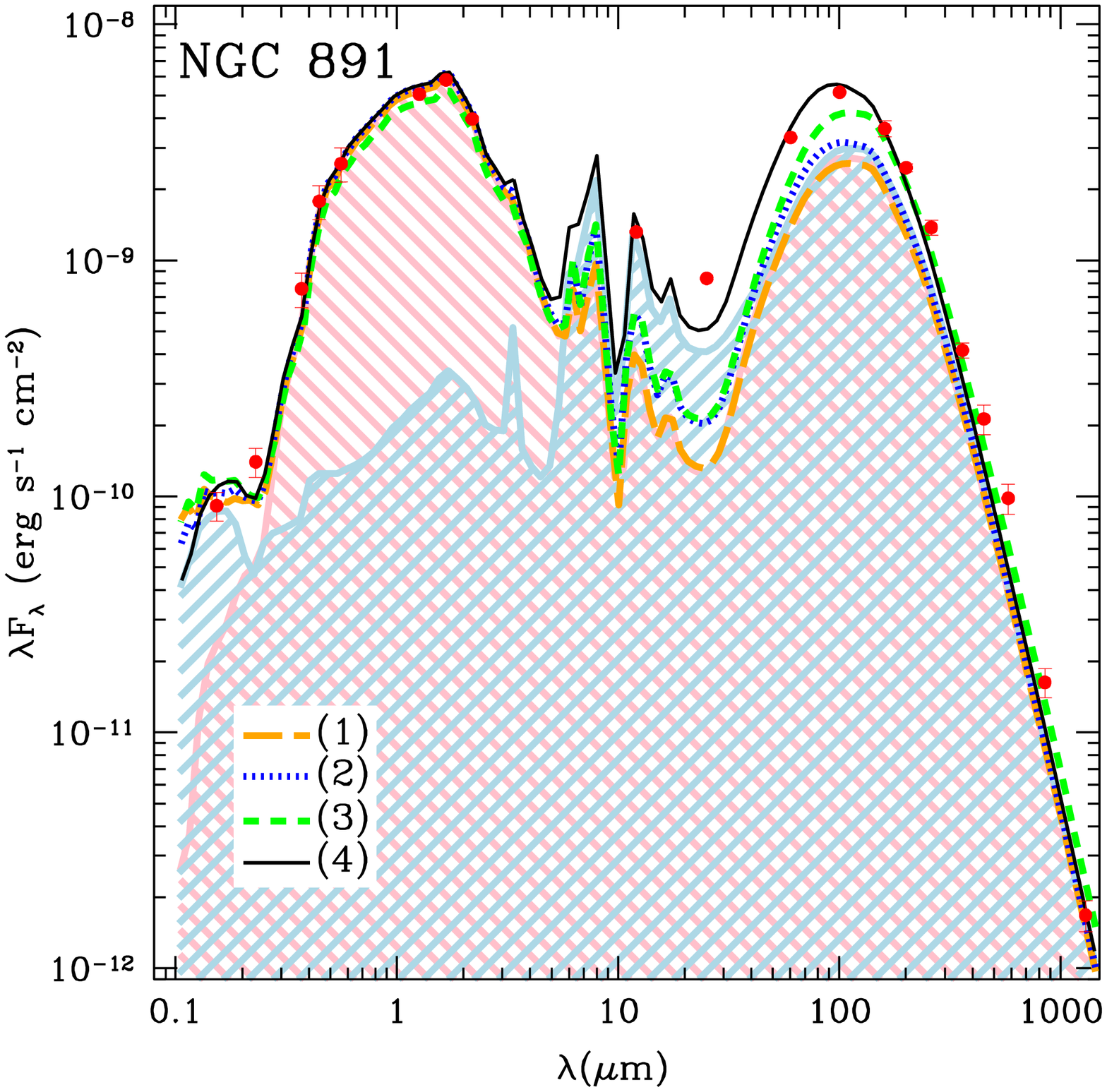}}
 \subfigure{\includegraphics[width=2.44in]{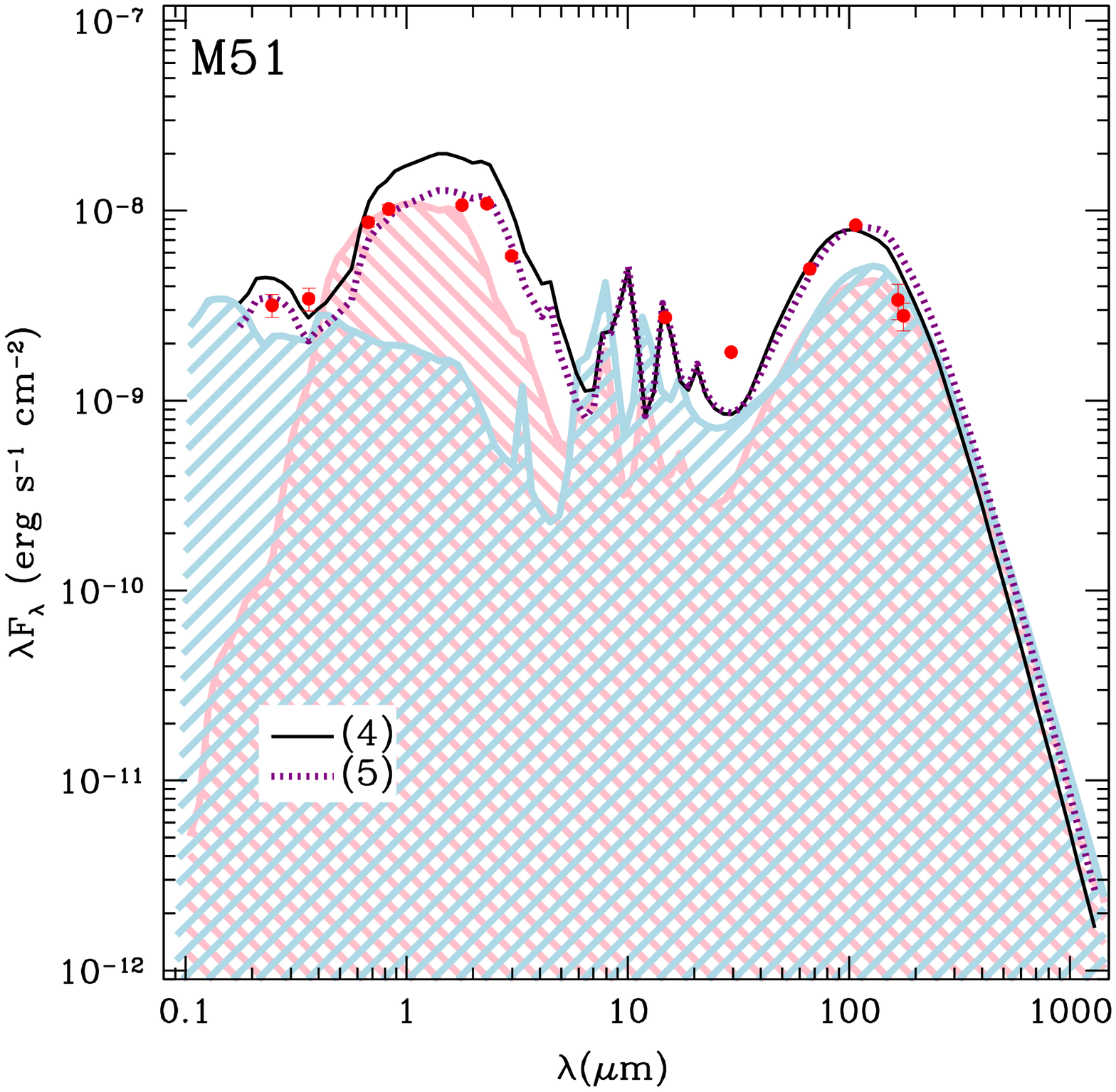}}
 \caption{Spatially integrated SEDs. Left: Two-disk models (lines) compared to data from NGC 891 (red points, from \cite[B08]{Bianchi08}). Pink and blue shading denote contributions to Model 4's SED from old and young stars, respectively. Right: Face-on two-disk models compared to data for M51 (from the {\it NASA/IPAC Extragalactic Database}).}
   \label{fig1}
\end{centering}
\end{figure}

Shown in Figure \ref{fig1}, restricting the young stellar emission to areas of higher dust density (Model 4) allows us to dramatically increase the luminosity of the second disk without increasing the dust mass. This density threshold appears to be unique for any combination of disk luminosities and dust masses --- higher thresholds push the dust peak too far into the blue, while at lower thresholds the model cannot simultaneously reproduce the UV, FIR, and 12 $\mu$m PAH feature. While this model still predicts a low 24 $\mu$m flux, high-resolution simulations of individual dust clouds indicate that this discrepancy is due to the unresolved nature of the dense clumps in our simulations (\cite[B08]{Bianchi08}). 
We investigated the viability of our models at other projected
inclinations by comparing them to M51 ($i\sim 20\deg$; Figure 1). For
simplicity we fix all geometric parameters (including clumping and
spirality) of the dust and stellar distributions as well as L$_2$, and allow
only M$_{\rm d}$ and L$_1$ to vary. A good match to the observed,
spatially integrated SED (except at 24$\mu$m, as explained above) is
found by increasing the dust mass by 75\% and halving the old-disk
luminosity (model 5). The FIR flux distributions indicate that M51 has
a dearth of cold dust compared to N891 despite M51's increased dust
mass, consistent with elevated star-formation induced by M51's ongoing
interaction. 

While the addition of spirality has little effect on the shape of the spatially-integrated SED, it does produce a gradient in the UV/blue light across the major axis of the model. This gradient is present in NGC 891 and is commonly considered to be due to the presence of a a grand design spiral pattern \cite{Kamphuis07}. This gradient presents an opportunity to better constrain the otherwise hidden thin disk of edge-ons, and can only be fit by models with non-axisymmetric structure like ours. 
\acknowledgements{Research was supported by NSF/AST1009471. With data obtained from MAST/STScI, and the NASA/IPAC Extragalactic Database (NED). We thank M. Wolff for help installing HYPERION; C. Howk, B. Benjamin, and B. Whitney for useful discussions.}


\begin{thebibliography}{}
\bibitem[Bianchi (2008)]{Bianchi08} Bianchi, S.\ 2008, \textit{A\&A}, 490, 461 
\bibitem[Fioc \& Rocca-Volmerange (1997)]{Fioc97} Fioc, M., \& Rocca-Volmerange, B.\ 1997, \textit{A\&A}, 326, 950 
\bibitem[Kamphuis \etal\ (2007)]{Kamphuis07} Kamphuis, P., et al.\ 2007, \textit{A\&A}, 471, L1 
\bibitem[Popescu \etal\ (2000)]{Popescu00} Popescu, C.~C., Misiriotis, A., Kylafis, N.~D., Tuffs, R.~J., \& Fischera, J.\ 2000, \textit{A\&A}, 362, 138 
\bibitem[Popescu \etal\ (2011)]{Popescu11} Popescu, C.~C., Tuffs, R.~J., Dopita, M.~A., et al.\ 2011, \textit{A\&A}, 527, A109
\bibitem[Robitaille(2011)]{Robitaille11} Robitaille, T.~P.\ 2011, \textit{A\&A}, 536, A79 
\bibitem[Schechtman-Rook et al.(2012)]{Schechtman-Rook12} Schechtman-Rook, A., Bershady, M.~A., \& Wood, K.\ 2012, \textit{ApJ}, 746, 70 
\bibitem[Xilouris \etal\ (1999)]{Xilouris99}
Xilouris, E. M., et al. 1999, \textit{A\&A}, 344, 868
\end{thebibliography}
\end{document}